\documentclass[aps,pra,twocolumn,superscriptaddress]{revtex4-1}
\usepackage[utf8]{inputenc}
\usepackage[english]{babel}
\usepackage{float}
\usepackage{amsfonts,amsmath,amssymb}
\usepackage{bm}
\usepackage{graphicx}
\usepackage{color}

\newcommand{\D}{\mathrm{d}}
\newcommand{\CP}{\mathrm{CP}}
\newcommand{\dB}{\mathrm{dB}}

\newcommand{\g}{\mathrm{g}}
\newcommand{\num}{\mathrm{num}}
\newcommand{\ana}{\mathrm{ana}}
\newcommand{\zer}{\mathrm{0}}
\newcommand{\one}{\mathrm{1}}

\newcommand{\tz}{\tilde{z}}
\newcommand{\tF}{\tilde{F}}
\newcommand{\tpsi}{\tilde{\psi}}

\newcommand{\bz}{\bm{z}}

\newcommand{\bF}{\bm{F}}
\newcommand{\bE}{\bm{E}}
\newcommand{\bV}{\bm{V}}
\newcommand{\bpsi}{\pmb{\psi}}

\newcommand{\cE}{\mathcal{E}}

\let\Re\oldRe
\let\Im\oldIm
\DeclareMathOperator{\Re}{Re}
\DeclareMathOperator{\Im}{Im}
\DeclareMathOperator{\Ai}{Ai}
\DeclareMathOperator{\Bi}{Bi}
\DeclareMathOperator{\Ci}{Ci}

\begin{document}

\title{Casimir-Polder shifts on quantum levitation states}

\author{P.-P. Cr\'epin} \email[]{pierre-philippe.crepin@lkb.upmc.fr}
\affiliation{Laboratoire Kastler Brossel, UPMC-Sorbonne
Universit\'es, CNRS, ENS-PSL Research University, Coll\`ege de
France, Jussieu case 74, F-75252 Paris, France.}
\author{G. Dufour} \email[]{gabriel.dufour@physik.uni-freiburg.de}
\affiliation{Laboratoire Kastler Brossel, UPMC-Sorbonne
Universit\'es, CNRS, ENS-PSL Research University, Coll\`ege de
France, Jussieu case 74, F-75252 Paris, France.}
\affiliation{Physikalisches Institut, Albert-Ludwigs-Universit\"{a}t Freiburg, 
Hermann-Herder-Stra{\ss}e 3, D-79104, Freiburg, Germany}
\author{R. Gu\'erout} \email[]{romain.guerout@lkb.upmc.fr}
\affiliation{Laboratoire Kastler Brossel, UPMC-Sorbonne
Universit\'es, CNRS, ENS-PSL Research University, Coll\`ege de
France, Jussieu case 74, F-75252 Paris, France.}
\author{A. Lambrecht} \email[]{astrid.lambrecht@lkb.upmc.fr}
\affiliation{Laboratoire Kastler Brossel, UPMC-Sorbonne
Universit\'es, CNRS, ENS-PSL Research University, Coll\`ege de
France, Jussieu case 74, F-75252 Paris, France.}
\author{S. Reynaud} \email[]{serge.reynaud@lkb.upmc.fr}
\affiliation{Laboratoire Kastler Brossel, UPMC-Sorbonne
Universit\'es, CNRS, ENS-PSL Research University, Coll\`ege de
France, Jussieu case 74, F-75252 Paris, France.}

\date{\today}

\begin{abstract}
An ultracold atom above a horizontal mirror experiences quantum reflection from the attractive Casimir-Polder interaction, which holds it against gravity and leads to quantum levitation states. We analyze this system by using a Liouville transformation of the Schr\"odinger equation and a Langer coordinate adapted to problems with a classical turning point. Reflection on the Casimir-Polder attractive well is replaced by reflection on a repulsive wall and the problem is then viewed as an ultracold atom trapped inside a cavity with gravity and Casimir-Polder potentials acting respectively as top and bottom mirrors. We calculate numerically Casimir-Polder shifts of the energies of the cavity resonances and propose a new approximate treatment which is precise enough to discuss spectroscopy experiments aiming at tests of the weak equivalence principle on antihydrogen. We also discuss the lifetimes by calculating complex energies associated with cavity resonances.
\end{abstract}

\maketitle

\section{Introduction}
\label{sec:intro}

Gravitational confinement of particles above a horizontal reflective surface is a common classical process.
In quantum physics, this process leads to the existence of quantum levitation states for ultracold matter waves.
Such gravitationally bound quantum states have been observed with ultracold neutrons~\cite{Nesvizhevsky2002nature,Nesvizhevsky2003,Nesvizhevsky2005}.
Though atomic mirrors have been realized using inhomogeneous electric or magnetic fields~\cite{Kasevich1990,Aminoff1993,Roach1995,Landragin1996,Henkel1996,Sidorov1996,Bongs1999,Rosenblum2014}, gravitationally bound quantum states of atoms remain to be observed.

At the low energies required to reach the quantum regime, quantum levitation states can be built up on quantum reflection above the Casimir-Polder (CP) tail created by the surface. Classically, the attractive force would be expected to accelerate the atom towards the surface, not to reflect it. In quantum physics, the matter wave can be efficiently reflected, due to the rapid variation of the CP potential at the approach of the surface 
(\cite{Lennard-Jones1936III,Lennard-Jones1936IV,Berry1972,Carraro1998,%
Shimizu2001,Friedrich2004a,Pasquini2004,Pasquini2006,Dufour2013qrefl,Dufour2013porous} and references therein). It should therefore be possible to trap atoms in quantum levitation states above a horizontal mirror, with gravity pulling them downwards and quantum reflection balancing their free fall \cite{Jurisch2006,Madronero2007}.

These quantum levitation states can play a useful role in the emerging investigation of gravitational properties of antimatter. It has recently be proposed to test the weak equivalence principle with antihydrogen by timing its free fall from a height prescribed by a trapping device (\cite{Debu2012,Indelicato2014,Perez2015,Dufour2015ahep} and references therein). The precision of this test, of the order of 1\% for the timing experiment, could be improved by using the  gravitational quantum states of antihydrogen above a material surface~\cite{Voronin2005jpb,Voronin2005pra}. 
The basic idea is that the properties of these states are essentially determined by gravity so that spectroscopic techniques can measure accurately the free fall acceleration for antihydrogen~\cite{Voronin2011,Voronin2014ijmp,Voronin2016jpb}. As these properties are affected by the proximity of the surface, an accurate determination of the acceleration requires a precise evaluation of Casimir-Polder shifts on quantum levitation states.

Quantum reflection becomes more efficient for weaker CP potential, due to the fact that it thus occurs closer to the surface, with the Casimir-Polder potential varying more and more rapidly there \cite{Dufour2013qrefl,Dufour2013porous}. This counterintuitive property has a nice interpretation in terms of Liouville transformations~\cite{Liouville1837,Olver1997} of the Schr\"odinger equation, which change the potential landscape while preserving the scattering amplitudes. When the classical action is used as a new coordinate, quantum reflection from the attractive CP well is transformed into ordinary reflection on a repulsive wall~\cite{Dufour2015epl,Dufour2015jpb}. Such a coordinate is not suitable for the study of quantum levitation states, as it cannot handle the connection problem~\cite{Wentzel1926,Kramers1926,Brillouin1926} at the turning point corresponding to the maximum altitude reached classically. We will solve this problem by using a better adapted coordinate system proposed by Langer~\cite{Langer1931,Langer1937}. 

In section~\ref{sec:bouncer}, we recall results for a quantum particle bouncing on an infinitely high and steep potential step, which will be useful as a reference for our discussions. We introduce Liouville transformations and Langer coordinates in section~\ref{sec:liouville}. We calculate the properties of the quasi-stationary states of the quantum bouncer in section~\ref{sec:cavity}, with the transformed physical picture corresponding to a cavity built up with two mirrors, a partly reflective one associated with quantum reflection and a perfectly reflecting one due to gravity. In section~\ref{sec:cpshifts}, we present an approximate description of these properties and show that it should be sufficient for the proposed spectroscopic tests of the weak equivalence principle with antihydrogen \cite{Voronin2011,Voronin2014ijmp,Voronin2016jpb}.

\section{Quantum bouncers}
\label{sec:bouncer}

We consider a particle of mass $m$ and energy $E$ in the Earth gravity field $g$ above a perfectly plane and horizontal mirror. The potential $V$ depends only on the altitude $z$ of the particle above the mirror sitting at $z=0$. The wavefunction $\psi(z)$ obeys the one-dimensional Schr\"odinger equation
\begin{align}\label{schrod}
-\frac{\hbar^2}{2m}\frac{\D^2\psi }{\D z^2}(z) +  V(z)\psi(z)=E \psi(z) ~.
\end{align}

Far away from the material surface, the free solution of Schr\"odinger equation in the linear potential $mgz$ is given by the Airy function $\Ai$  \cite{Olver2010}
\begin{align}\label{unconstrained}
&\psi(z)=a\Ai\left( \frac{z-z_t}{\ell_\g} \right)=a\Ai\left( \frac{z}{\ell_\g}-\frac{E}{\epsilon_\g} \right) ~,\\
&z_t\equiv\frac{E}{mg} ~.
\end{align}
Here $z_t$ corresponds to the classical turning point while $\ell_\g$ and $\epsilon_\g$ are length and energy scales associated with quantum effects in the gravity field $g$
\begin{align}
\ell_\g&=\left( \frac{\hbar^2}{2m^2 g} \right)^{1/3}\approx 5.87~\mu\text{m}~,\label{grav-length}\\
\epsilon_\g&=mg\ell_g\approx 0.602~\mathrm{peV}~.\label{grav-energy}
\end{align} 
Throughout the paper, numerical values correspond to a hydrogen atom falling in the standard Earth gravity field $g\approx 9.81~$m.s$^{-2}$. They would have to be changed if the acceleration were different for antihydrogen.
Note that the Airy function $\Bi$, which could appear in the unconstrained solution \eqref{unconstrained}, has been discarded as its asymptotic behavior above the classical turning point describes an unphysical exponentially growing wave.

In the ideal quantum bouncer model, the atom is perfectly reflected by an infinitely steep wall at $z=0$.
This model, which is suitable  for neutrons bouncing off the Fermi potential step resulting from the strong interaction of the wavepacket with nuclei in the mirror \cite{Nesvizhevsky2015}, enforces the boundary condition $\psi(0)=0$. It leads to energy levels $E_n^\zer$ of the ideal quantum bouncer determined by the zeros of the function $\Ai$ 
(the superscript $\zer$ indicates that CP shifts are not taken into account here)
\begin{align}\label{energies-ideal}
 & E_n^\zer=\lambda_n \epsilon_\g~, & & \Ai(-\lambda_n)= 0~, && n=1,2,\dots  
\end{align}
The function $\Ai$ has a countable infinity of zeros, all with negative values on the real line.
Numerical values of the first ones are given in Table 9.9.1 of the Digital Library of Mathematical Functions (see \cite{Olver2010}). Asymptotic expansions of their values are also given in \S9.9(iv) of the same resource, with a very good accuracy for $n$ larger than 10.

In this work, we study the case of atoms bouncing off the surface of the mirror due to quantum reflection on the CP potential $V_\CP(z)$. This corresponds to solutions of Schr\"odinger equation \eqref{schrod} with the potential
\begin{align}
\label{fullpotential}
 V(z)=mgz+V_\CP(z)~.
\end{align}
We also suppose that atoms are absorbed when touching the surface, which corresponds to the physical boundary condition for antihydrogen annihilated when reaching contact with matter. For an experiment performed with atoms, this assumption should be justified or replaced by proper physical boundary conditions at contact with the surface.
We follow the method exposed in \cite{Dufour2013qrefl} to calculate the exact potentials corresponding to antihydrogen atom above a mirror perfectly reflecting the electromagnetic field, a silicon bulk or a silica bulk.

The potential, attractive at all distances, behaves as $V_\CP(z)\simeq-C_3/z^3$ near the surface and $V_\CP(z)\simeq-C_4/z^4$ far from the surface \cite{Casimir1948physrev}. The long-range tail is characterized by the following length and energy scales
\begin{align}
\ell_\CP&=\frac{\sqrt{2mC_4}}{\hbar} \approx 27.5~\mathrm{nm}~,\\
\epsilon_\CP&=\frac{C_4}{\ell_\CP^4} \approx 27.4~\mathrm{neV}~.
\end{align}
Here the numerical values correspond to antihydrogen above a mirror perfectly reflecting electromagnetic fields.
These length and energy scales are respectively much smaller and much larger than those (\ref{grav-length}-\ref{grav-energy}) associated with the gravitational potential. An approximate solution of the problem can thus be found by decoupling the effects of gravity and CP interactions. The solutions of the Schr\"odinger equation are given by the unconstrained solution \eqref{unconstrained} when the atom is far away from the surface, whereas the scattering on the CP potential modifies the boundary condition at $z$ of the order of $\ell_\CP$.

For the lowest quantum states with not too large values of $n$ ($\lambda_n\epsilon_\g\ll\epsilon_\CP$), the scattering amplitudes are mainly given by the scattering length. It follows that the energies are shifted by a quantity $mga$ resulting from the complex phase shift experienced by the atom upon reflection on the CP tail \cite{Voronin2005jpb,Voronin2005pra}
(the calligraphic $\cE$ signals that those energies are complex while the superscript $\one$ indicates that CP shifts are calculated in a first approximation, to be improved in the following)
\begin{align}\label{constant-shift}
\cE_n^\one=\lambda_n \epsilon_\g+mga~.
\end{align}
The imaginary part of the complex shift is related to the lifetime of the quasi-stationary states 
(more detailed discussions in section~\ref{sec:complex}).

Within the approximation \eqref{constant-shift}, called the \emph{scattering length approximation} in the following,
the transition frequencies \cite{Kreuz2009,Jenke2011,Baessler2015} between quantum states are independent of the atom-surface interaction 
\begin{align}\label{constant-transition}
\omega_{mn}^\one= \frac{\cE_n^\one-\cE_m^\one}{\hbar}= \frac{E_n^\zer-E_m^\zer}{\hbar} = \omega_{mn}^\zer~.
\end{align}
Therefore spectroscopy experiments on transitions between quantum states give access to the value of $\epsilon_\g$, that is also $g$,  while being unaffected by the details of the interaction with the surface. This is the key idea opening perspectives for testing the free fall on antihydrogen through accurate frequency measurements.  In the sequel of this paper, we perform an exact treatment of the full potential including the effects of gravity and CP interaction, which will allow us to assess the accuracy of the approximation \eqref{constant-shift}. We also give improved numerical and analytical results sufficient for discussing the proposed spectroscopic tests of free fall
\cite{Voronin2011,Voronin2014ijmp,Voronin2016jpb}.

\section{Liouville transformations}
\label{sec:liouville}

Liouville transformations are a group of transformations which preserve the scattering amplitudes obtained by solving the Schr\"odinger equation, while changing the semiclassical landscape of the problem. 

The Schr\"odinger equation can be written in the following compact form
\begin{align}\label{schrodF}
& \psi''(z)+F(z)\psi(z)=0~,\\
& F(z)=k_\dB(z)^2=\frac{2m}{\hbar^2}\left( E-V(z) \right)~,
\end{align}
where $F(z)$ is the square of the de Broglie wavevector $k_\dB$. 
A Liouville transformation of equation \eqref{schrodF} consists in  a coordinate change $z\to\tz$ associated with a rescaling $\psi\to\tpsi$ of the wave-function \cite{Liouville1837,Olver1997} 
\begin{align}\label{tpsi}
\tpsi(\tz)=\sqrt{\tz'(z)}\psi(z) 
\end{align}
The coordinate change maps the physical $z-$domain into a $\tz-$domain with $\tz(z)$ a smooth monotonous function ($\tz'(z)>0$). 
The transformed wave function also obeys a Schr\"odinger equation with a transformed $F-$function
\begin{align}
\label{tschrod}
&\tpsi''(\tz) + \tF(\tz)  \tpsi(\tz)=0~,\\
\label{tF}
&\tF(\tz)=\frac{F(z)-\frac{1}{2}\{\tz,z\}}{\tz'(z)^2}~.
\end{align}
The curly braces denote the Schwarzian derivative of the coordinate transformation 
\begin{equation}
\{\tz,z\}=\frac{\tz'''(z)}{\tz'(z)}-\frac{3}{2}\frac{\tz''(z)^2}{\tz'(z)^2}~.
\end{equation}

Although the functions $\tF(\tz)$ and $F(z)$ can be radically different, thus corresponding to different semiclassical landscapes, the two Schr\"odinger equations share a number of properties. The Liouville transformations have the remarkable property of preserving the Wronskian of any two solutions and, consequently, the physical properties which can be expressed in terms of Wronskians. In particular, scattering amplitudes are determined by Wronskians of specific solutions of the Schr\"odinger equation and preserved by the Liouville transformation.

In \cite{Dufour2015epl,Dufour2015jpb}, a Liouville transformation was performed by using the classical action, or WKB phase, as the new coordinate well defined only when $F(z)$ is positive
\begin{align}
\tz(z)=\phi_\dB(z)=\int^z k_\dB(\zeta)\D \zeta=\int^z \sqrt{F(\zeta)}\D \zeta~.
\end{align}
This coordinate can only be used below the classical turning point and it cannot help to study the connection with the region above this point~\cite{Wentzel1926,Kramers1926,Brillouin1926}.
We now introduce a coordinate proposed by Langer~\cite{Langer1931,Langer1937} which is
well adapted to this problem, as it leads to a wavefunction regular at the crossing of the turning point. 

The turning point $z_t$ corresponds to $F(z_t)=0$ and $F'(z_t)<0$, with
$F(z)$ nearly linear in its neighborhood
\begin{align}
F(z)\underset{z \to z_t}{\simeq} -F'(z_t)(z_t-z)~.
\end{align}
The Langer coordinate $\bz$ is defined so that $\bF(\bz)$ shows the same linear behavior in the vicinity of $\bz_t=\bz(z_t)$ (boldfaces denote all quantities related to the Langer coordinate system)
\begin{align}\label{bF-tp1}
&\bF(\bz)\underset{\bz \to \bz_t}{\simeq} \bz_t-\bz~,\quad\bz_t=\frac{E}{\epsilon_\g}~.
\end{align}
We have partly used the freedom in the definition of the Langer coordinate $\bz$, by fixing  $\bz_t$ and $\bF'(\bz_t)=-1$. 

The change of coordinate $\bz(z)$ reduces to a linear function near the turning point
\begin{align}
\bz(z)& \underset{z \to z_t}{\simeq}\bz_t+ \left(  -F'(z_t)\right)^{1/3}(z-z_t)~,
\end{align}
and the Schwarzian derivative $\{\bz,z\}$ vanishes around the turning point, with equation \eqref{tF} reducing to
\begin{align}\label{bF-tp2}
\bF(\bz)\underset{\bz \to \bz_t}{\simeq} \frac{F(z)}{\bz'(z)^2}~.
\end{align}
We now fix the definition by requiring the right hand sides of equations \eqref{bF-tp1} and \eqref{bF-tp2} to be equal for all $z$
\begin{align}\label{def-langer}
&\bz'(z)=\sqrt{\frac{F(z)}{\bz_t-\bz}}~,\quad \text{for} \quad z\neq z_t~,\\
&\bz'(z_t)=\left(-F'(z_t) \right)^{1/3} ~.
\end{align}
Evaluating equation \eqref{tF} for the Langer coordinate yields 
\begin{align}\label{bF}
\bF(\bz)&=\bz_t- \bz\\
&-\frac{5}{16(\bz-\bz_t)^2}+ (\bz- \bz_t)Q(z)~,\notag
\end{align}
where $Q(z)$ is the so-called \emph{badlands} function \cite{Dufour2013qrefl} expressed in terms of the initial coordinate $z$
\begin{align}
Q(z)=\frac{F''(z)}{4 F(z)^2}-\frac{5 F'(z)^2}{16 F(z)^3}~.
\end{align}
Note that the two last terms in equation \eqref{bF} diverge at the turning point while their sum does not.
In regions where $\bF(\bz)\simeq (\bz_t- \bz) $, in particular around the turning point, the Schr\"odinger equation reduces to an Airy equatio, and the connection problem at the turning point is solved as in \eqref{unconstrained} by the Airy function $\Ai$
\begin{equation} \label{AUA}
\bpsi(\bz) \simeq a \Ai(\bz-\bz_t)~.
\end{equation}

\begin{figure}[t]
\centering
\includegraphics[width=0.45\textwidth]{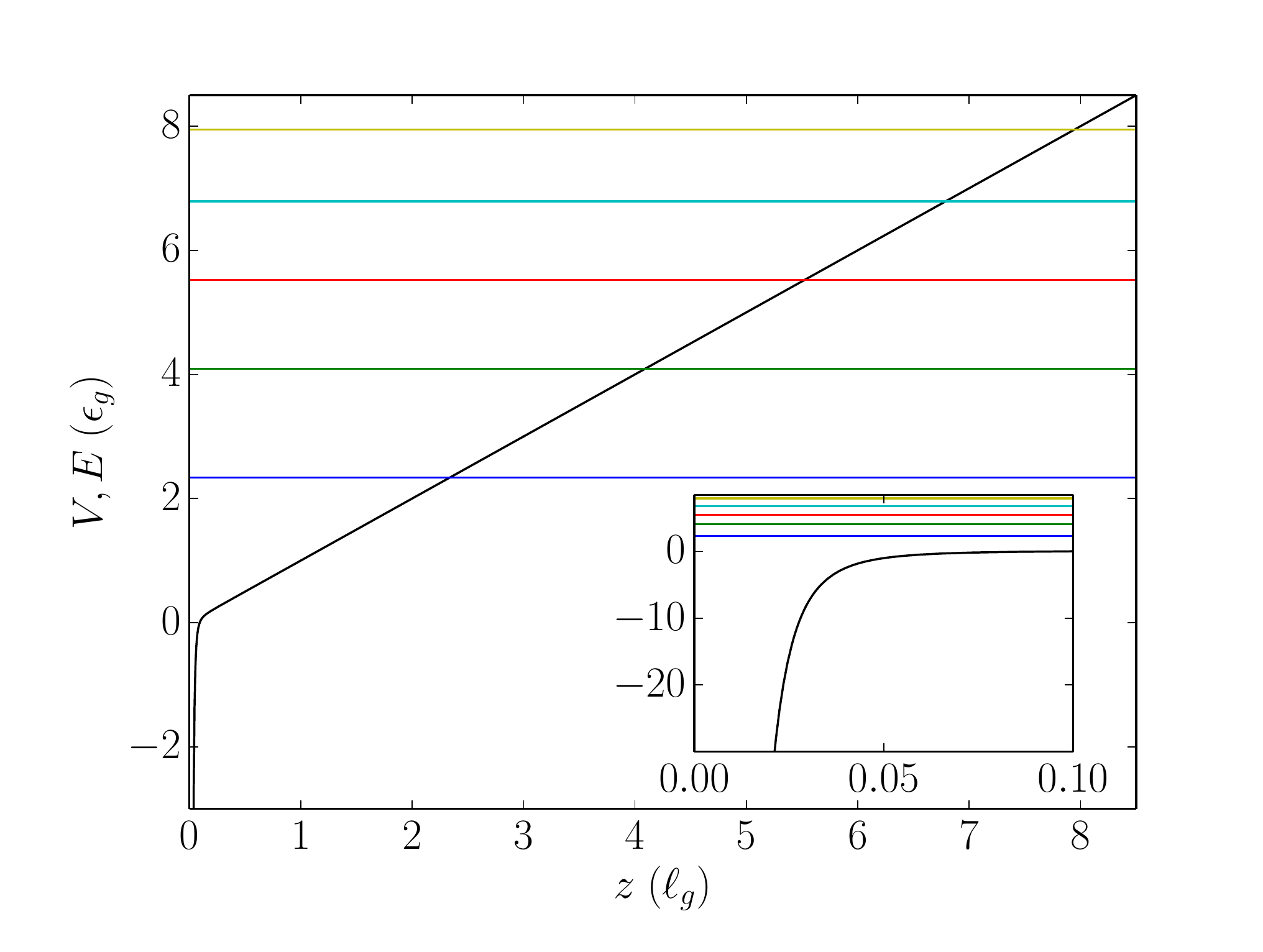}
\caption{\label{original-silica} The original problem for antihydrogen atom in the combined gravity and CP potentials $V(z)=mgz+V_\CP(z)$ (black curve) above a silica bulk. Horizontal lines correspond to energies chosen for the illustration as $E_n^\zer=\lambda_n\epsilon_\g,~n=1,...,5$  (blue, green, red, cyan and yellow lines, respectively from bottom to top line). A zoom on the potential well near the surface is shown in the inset. [Colors online]} 
\end{figure}

In the following, we give a complete solution of the Schr\"odinger equation \eqref{tschrod}, keeping all terms in the expression \eqref{bF} of $\bF$. 
The solution is therefore fully equivalent to the exact solution of the original Schr\"odinger equation \eqref{schrod}, illustrated on Figure \ref{original-silica} for antihydrogen atom above a silica bulk \cite{Dufour2013qrefl}. The horizontal lines are drawn for energies \eqref{energies-ideal} matching the states $n=1,...,5$ (blue, green, red, cyan and yellow lines, from bottom to top line)  of an ideal quantum bouncer. The exact energies, shifted with respect to \eqref{energies-ideal} due to the effect of the CP interaction, are calculated in the following.

Figure \ref{transformed-silica} shows the same problem as on Figure \ref{original-silica} now treated in the Langer coordinate system, with the coordinate $\bz$ spanning the whole real axis and the transformed $\bF-$function \eqref{bF} written in terms of a transformed energy $\bE$ and a transformed potential $\bV$
\begin{align}
&\bF(\bz)=\bE-\bV(\bz)~,\quad \bE=\bz_t=\frac{E}{\epsilon_\g} ~,\\
&\bV(\bz)=\bz-\bV_\CP(\bz) ~.
\end{align}
The potential $\bV$  is the sum of a linear gravity potential and an effective potential $\bV_\CP(\bz)$ producing quantum reflection.
In sharp contrast with the CP well on Figure \ref{original-silica}, the transformed potential $\bV_\CP$ now shows a high peak close to the surface. Its height is much larger than the energies of the lowest quantum states illustrated by the horizontal lines on Figure \ref{transformed-silica}, at $\bE_n^\zer=\lambda_n,~n=1,...,5$  (same color codes as on Figure \ref{original-silica}).

\section{Quantum levitation states}
\label{sec:cavity}

With quantum reflection understood as classically expected reflection on a repulsive wall, 
we get a new physical picture for quantum levitation states corresponding to matter waves trapped in a Fabry-Perot cavity.  The top mirror of the vertical cavity perfectly reflects matter waves due to gravity, while the bottom mirror partially reflects them due to quantum reflection. 
We interpret the properties of quantum levitation states in terms of cavity resonances, by performing calculations in analogy with the theory of optical Fabry-Perot cavities \cite{Lambrecht2006}. 

\begin{figure}[t]
\centering
\includegraphics[width=0.45\textwidth]{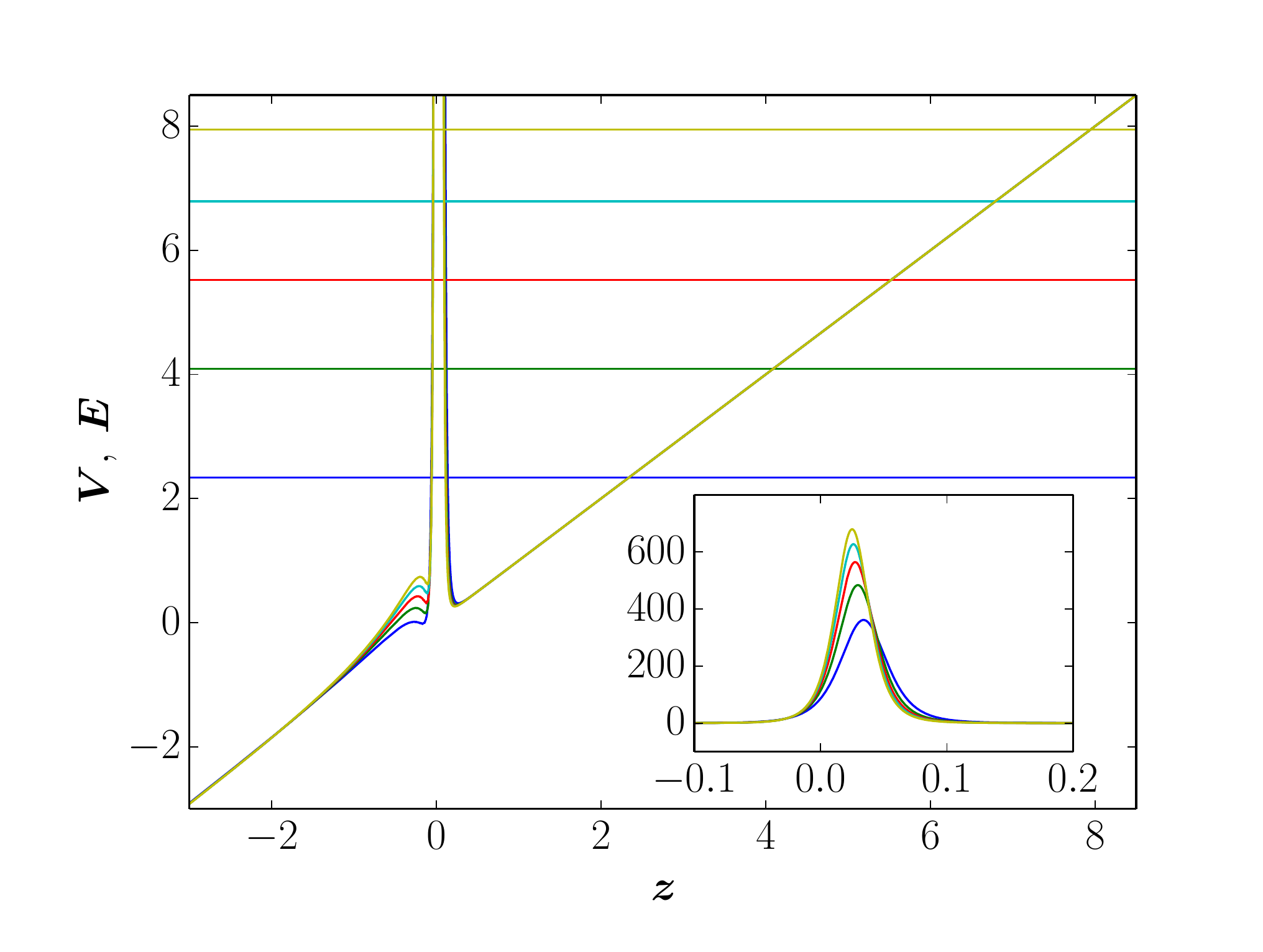}
\caption{\label{transformed-silica} The same problem as on figure \ref{original-silica} after a Liouville transformation to the Langer coordinate $\bz$, with energies chosen for the illustration as  $\bE_n^\zer=\lambda_n,~n=1,...,5$ and potential $\bV=\bz+\bV_\CP$. The color code is the same as in figure \ref{original-silica}. A zoom on the wall is shown in the inset. [Colors online]} 
\end{figure}

Around and above the top mirror, that is also around and above the turning point $\bz_t$, the solution of the Schr\"odinger equation is given by the Airy function \eqref{AUA}, that we rewrite as a linear superposition of upward and downward traveling waves $\Ci^+$ and $\Ci^-$  
\begin{align} \label{travelingwaves}
&\bpsi_m(\bz) = \frac{a_m}2  \left(\Ci^+(\bz-\bz_t)+\Ci^-(\bz-\bz_t)\right)~,\\
&\Ci^\pm(\bz)=\Ai(\bz)\pm i\Bi(\bz)~. 
\end{align}
The upward and downward waves have an equal amplitude for the reason already discussed for equation \eqref{unconstrained}, that is the absence in \eqref{travelingwaves} of the combination $\Bi$ corresponding to an exponentially growing wave above the turning point. This amplitude is denoted $a_m$ in \eqref{travelingwaves} as it depends on the number $m$ of bounces of the matter wave on the bottom mirror, as explained now. 

With the {ideal quantum bouncer} model, the ideal energy levels $E_n^\zer$ would be recovered by obtaining the stationary quantum solutions of \eqref{travelingwaves}. But the more general problem studied in this paper is not unitary since atoms transmitted through the bottom mirror are lost (antihydrogen going through the bottom mirror is annihilated when reaching contact with the matter plate). As a consequence, the quantum levitation states can only be obtained as quasi-stationary states, with the amplitude $a_m$ decreasing after each bounce, due to the losses. 
In analogy with the theory of optical Fabry-Perot cavities \cite{Lambrecht2006}, we introduce a factor describing the modification of the traveling waves after one cavity round trip 
\begin{align}\label{roundtripfactor}
&a_{m+1}= \rho  a_{m}~.
\end{align}

This round trip factor can be obtained by solving numerically the quantum reflection problem on the bottom mirror of the cavity (i.e. the peak in the potential drawn on Fig.\ref{transformed-silica}). Precisely, the Schr\"odinger equation \eqref{tschrod} is solved with appropriate boundary conditions far from this mirror~: above the mirror ($\bz\to\infty$), the downward traveling wave matches the component proportional to $a_m\Ci^-$ in \eqref{travelingwaves} while the upward traveling wave matches the component proportional to $a_{m+1}\Ci^+=\rho a_{m}\Ci^+$; below the mirror ($\bz\to-\infty$), the downward traveling wave is proportional to $b_m\Ci^-$, with $b_m$ related to $a_m$ by a transmission amplitude, whereas the upward traveling wave vanishes there.

This procedure produces a complex-valued function $\rho(E)$ of the energy $E$, that is also of the altitude $\bz_t$ of the turning point. Stationary quantum states would correspond to the condition $\rho=1$ which cannot be met in the presence of losses. But we may define energies $E_n$ of quasi-stationary states by requiring $\rho(E_n)$ to be a real number slightly smaller than unity. The value attained for $\rho(E_n)$ is related to the loss at each bounce, that is also the finesse of the cavity resonance. This relation will be discussed in more details below in \S\ref{sec:complex}.

\begin{figure}[t]
\centering
 \includegraphics[width=0.45\textwidth]{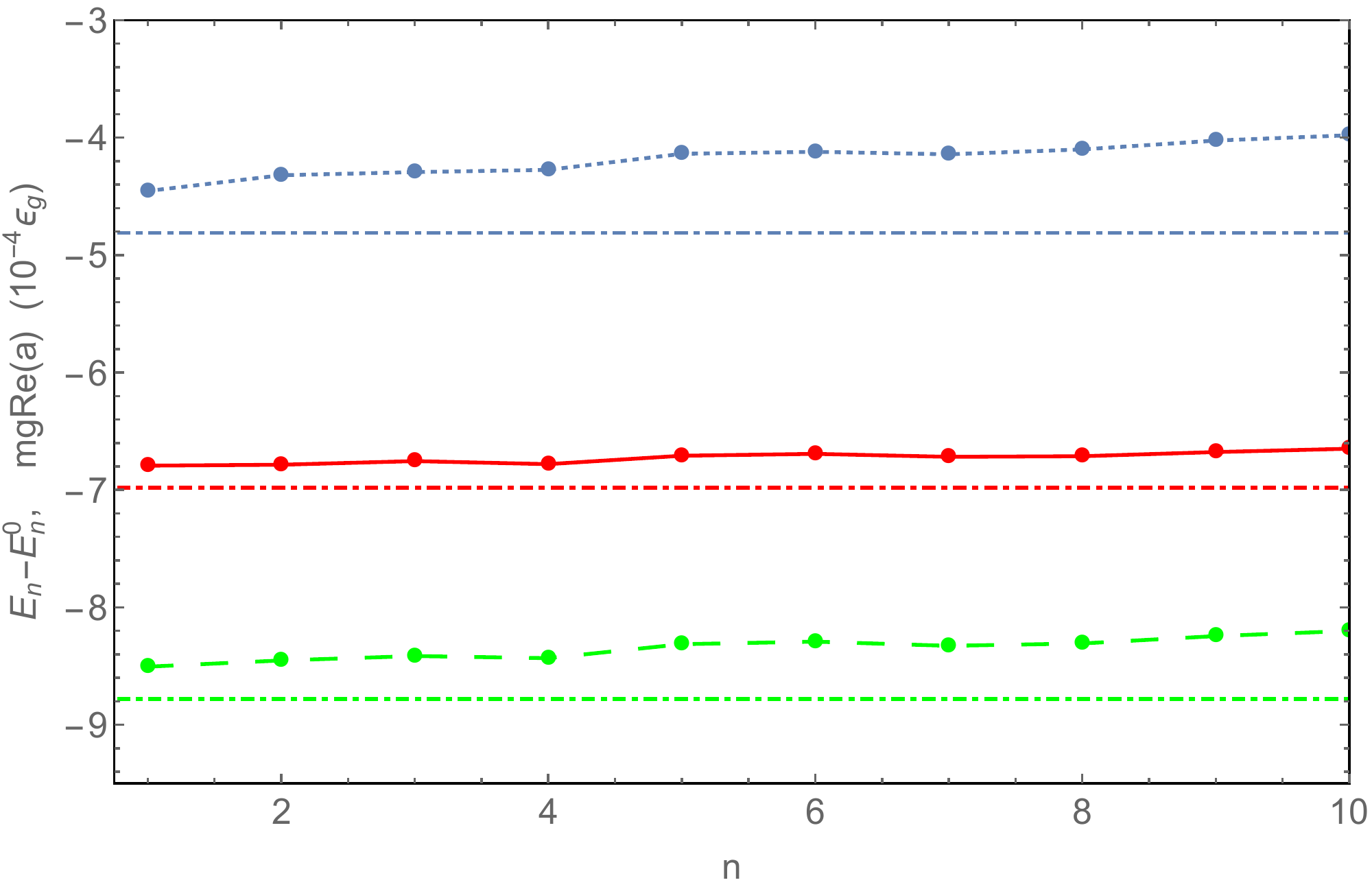}
\caption{\label{quasistates}  Energy shift $E_n-E_n^\zer$ for antihydrogen interacting with a perfect mirror (blue, top lines), a silicon bulk (green, bottom lines) or a silica bulk (red, middle lines), in units of $10^{-4}\epsilon_\g$. The shift corresponding to the real part of $mga$ is represented by the horizontal lines. [Colors online] }
\end{figure}

The energy shifts due to the CP effect are seen as the non vanishing differences $E_n-E_n^\zer$ between the numerical values $E_n$ obtained here and the expressions $E_n^\zer=\lambda_n\epsilon_\g$ calculated for the ideal quantum bouncer. They are shown on Figure \ref{quasistates} for the ten first resonances $n=1,2,\ldots,10$. The shifts for a given mirror are close to the constant value $mg\Re(a)$ predicted by the scattering length approximation \eqref{constant-shift} discussed above. The latter is confirmed as a first approximation of the numerical results, with an accuracy of the order of $10^{-4}$. A better analytical approximation, corresponding to an improved accuracy for the analysis of spectroscopy tests of free fall, is described in the next section.

\section{Casimir-Polder shifts}
\label{sec:cpshifts}

The round-trip factor $\rho$ is a scattering amplitude which can be evaluated in terms of Wronskians of solutions and, therefore, can be calculated in the initial or transformed coordinate systems equivalently. We now design an analytical approximation of this factor, built up on the optical analogy discussed in the preceding section.

The round trip factor $\rho$ is approximated as the product of two factors, the quantum reflection amplitude $r$ on the CP tail, and a propagation phase factor deduced from the phase $\theta$ of the Airy functions
\begin{align}\label{defrho}
&\qquad \rho \simeq  -r e^{2i\theta\left(-\bz_t\right)}~, \\
&\tan\theta(x) = \frac{\Ai(x)}{\Bi(x)}~,
&e^{2i\theta(x)}=-\frac{\Ci^-(x)}{\Ci^+(x)}~.
\label{Airyphase}
\end{align}
The determination of the solution in \eqref{Airyphase} is such that $\theta(0)=\pi/6$ with $\theta(x)$ a continuous function  \cite{Olver2010}.
After the discussions in the preceding section, the resonance energies $E_n$, with $\rho(E_n)$ a real number slightly smaller than unity, are the solutions of the equation
\begin{align}
2\theta\left(-\bz_t\right)+\arg \left(-r\right)=2n\pi~.
\label{resonenergy}
\end{align}
$\theta$ can be evaluated in the initial or Langer coordinate system, and it depends on the single parameter  
\begin{align}
\bz_t=\bE=\frac {E}{\epsilon_\g} = \frac {z_t}{\ell_\g} ~.
\end{align}
The argument of the complex amplitude $\left(-r\right)$ depends on the energy $E$ or on the equivalent wavevector $k$.

For a perfect quantum reflection $r=-1$, the equation \eqref{resonenergy} would give the energy levels $E_n^\zer$ as the zeros  $(-\lambda_n)$ of the Airy function $\Ai$ also obey $\theta(-\lambda_n)=n\pi$. With $\arg \left(-r\right)$ replaced by its scattering length approximation $\left( -2 k \Re(a) \right)$, and the equation \eqref{resonenergy} solved perturbatively in the small parameter $k a$, the energies $E_n$ are recovered as the real parts of $\cE_n^\one$ given in \eqref{constant-shift}. In the following, we use the \emph{effective-range approximation} \cite{Blatt1949,Bethe1949} which is much more accurate than the scattering length approximation. More precisely, we use the extension of effective-range approximation well suited for potentials having a long-range CP tail \cite{Spruch1960,OMalley1961,Macri2002,Arnecke2006}.

The reflection coefficient is thus written as a function of the wavevector $k$ and a complex length $\mathcal{A}(k)$ 
\begin{align}
r=-\frac{1-ik\mathcal{A}(k)}{1+ik\mathcal{A}(k)}\;,\quad \hbar k\equiv \sqrt{2mE} ~.
\label{effectiverange} 
\end{align}%
The limit $\mathcal{A}(0)$ is the scattering length $a$, but $\mathcal{A}(k)$  is now a function of $k$.
For the model potential exactly described by the homogeneous form $V_4\equiv-C_{4}/z^{4}$, the function $k\mathcal{A}$ is a known universal function \cite{Vogt1954,Gao2013} of the dimensionless parameter $k\ell $. 
Its expansion at low values of $k$ is deduced as (\cite{Spruch1960,OMalley1961} or eq.53 in \cite{Arnecke2006}) 
\begin{align} 
&k\mathcal{A} =-ik\ell ~\alpha \left( k\ell \right) ~, \quad
\ell=\frac{\sqrt{2mC_4}}{\hbar} ~, \\
&\alpha (K)  = \alpha _0+i\frac{\pi }{3} K
+\left(\alpha _2 +\frac{4}{3}\alpha _0 \ln K \right) K^2 ~,
\label{expansion} 
\end{align}
with known coefficients ($\gamma$ is the Euler constant)
\begin{align}  \label{expansionV4} 
&\alpha_0 = 1 ~, \quad 
\alpha_2 = \frac{8}{3}(\gamma +\ln 2)-\frac{28}{9}-\frac{2\pi }{3}i ~.
\end{align}

The exact potentials describing CP interaction of an antihydrogen atom with perfectly
reflecting surfaces, silicon or silica bulk, contain long-range tails $V_{4}$, but are not reducible
to these tails. Here, we use the numerical values obtained for $r(k)$ in \cite{Dufour2015jpb}, deduce $k\mathcal{A}$ by inverting \eqref{effectiverange}, and fit coefficients $\alpha _0$ and $\alpha _2$ in \eqref{expansion} to match its low energy expansion.
The coefficients obtained in this manner, given in Table 1, differ from those given above for the homogeneous potential $V_{4}$ and depend on the surface.
The difference is larger for weaker CP potentials, as could be expected from discussions in \cite{Dufour2015jpb}. 
Quantum reflection indeed occurs closer to the surface for weaker potentials, and the homogeneous form $V_{4}$  is not a good approximation of the real $V$ there.

\begin{table}[h]
\centering
\begin{tabular}{|c|c|c|c|}
\hline  \;             & Perfect mirror      &  Silicon bulk       &  Silica bulk     \\
\hline $\ell$         & 520.06 $a_0$      & 429.82 $a_0$    & 321.31 $a_0$         \\
\hline $\alpha_0$ & \, 1.0468$-$0.1028$i$ \, & \, 1.0149$-$0.2271$i$ \, & \, 0.8504$-$0.2414$i$ \, \\
\hline $\alpha_2$ & 0.17$-$2.06$i$        & 0.09$-$2.09$i$     & 0.70$-$4.8$i$   \\ 
\hline
\end{tabular}
\caption{Coefficients of the expansion of $\alpha$ obtained from a fit of the numerically calculated values of $r(k)$. $\ell$ is expressed in atomic units, with $a_0$ the Bohr radius. }
\end{table}

Within the effective-range approximation, the resonance energies are given as solutions of the equation
\begin{align}
\label{qsenergy}
&\theta\left(-\bE_n\right)-\Re\left(\arctan \left(K_n\alpha(K_n)\right)\right)=n\pi ~,\\
&\bE_n\equiv\frac{E_n}{\epsilon_\g} ~,\quad K_n\equiv k_n\ell = \frac{\sqrt{2mE_n}\ell}\hbar ~.
\end{align}
In \eqref{qsenergy} $\alpha$ is  defined by the expansion \eqref{expansion} with the coefficients in Table 1.
These coefficients have been obtained by fitting $r(k)$ on the interval of energies from 0 to 500$\epsilon_\g$.
The upper bound of the interval was chosen small enough for \eqref{expansion} to remain a good approximation of the function and, at the same time, large enough to keep a low numerical noise in the fit.
We have checked that the truncated numerical values given in the table (with a precision for the coefficient $\alpha_p$  decreasing with $p$) are sufficient to reproduce the variation of $r(k)$ for the purpose of our calculations, with errors in the evaluation of energies smaller than a few 10$^{-6}\epsilon_\g$ on the interval $0<E<500 \epsilon_\g$.

\begin{figure}[t]
\centering
   \includegraphics[width=0.45\textwidth]{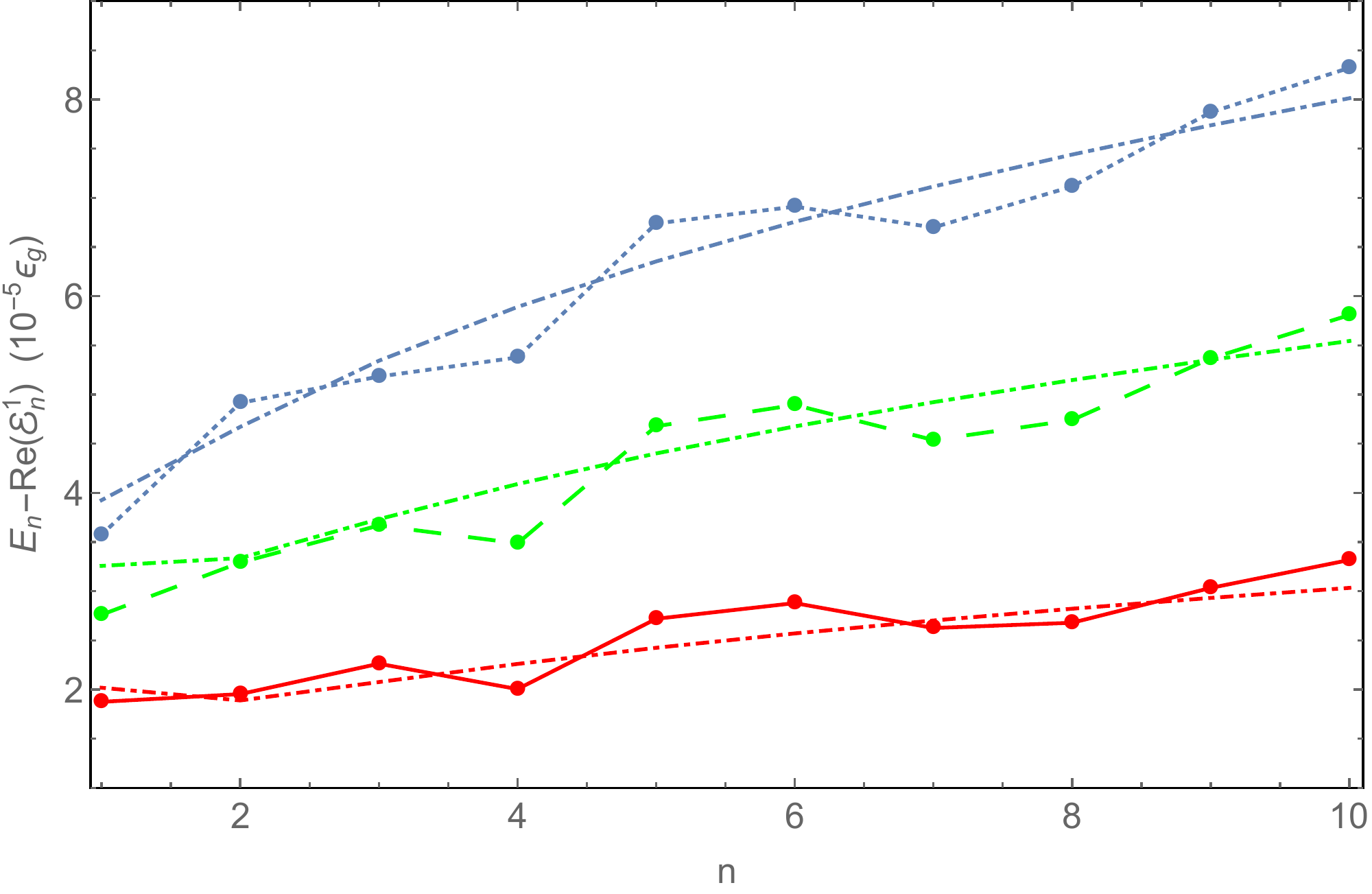}
   \includegraphics[width=0.45\textwidth]{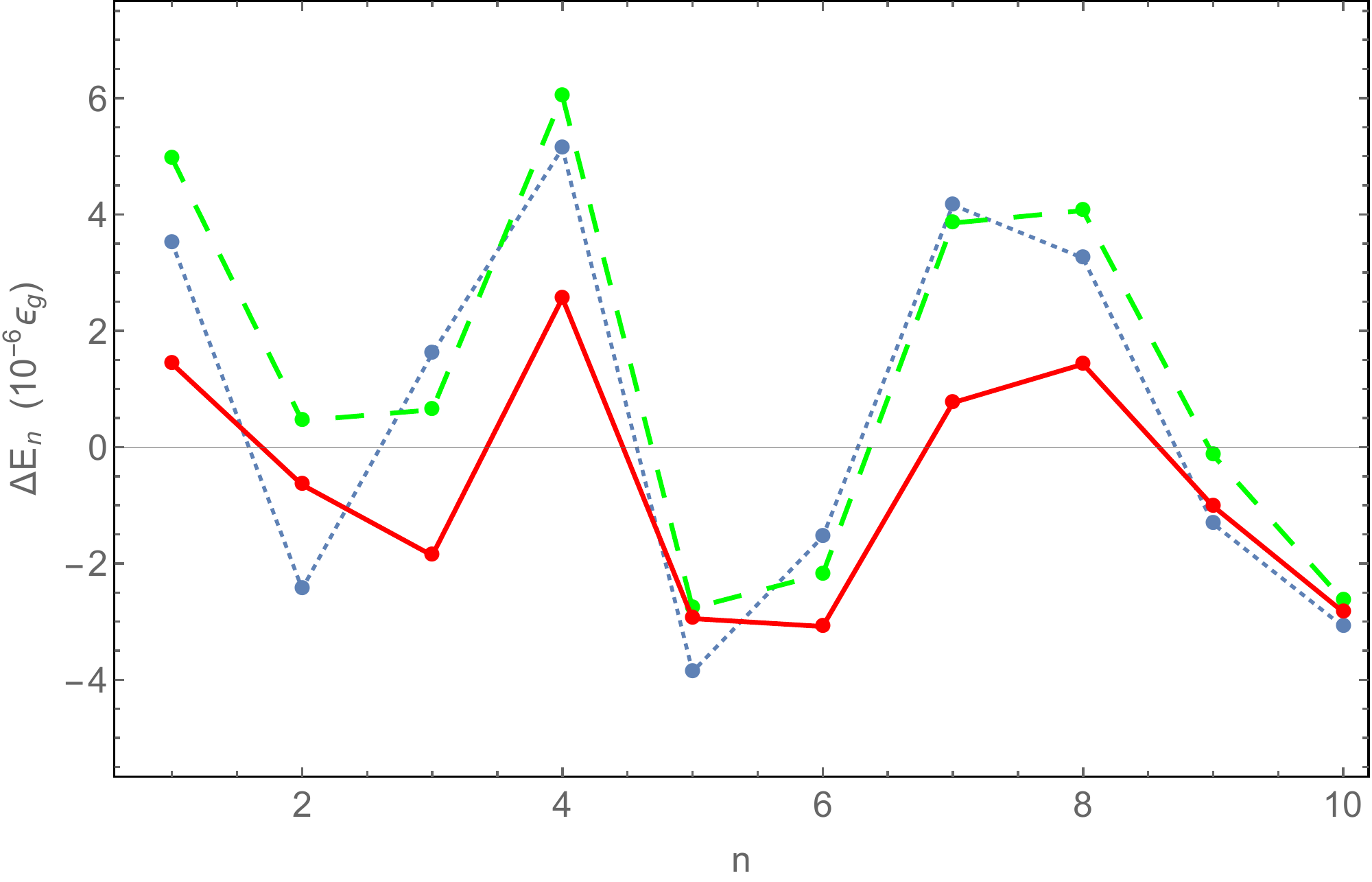}
\caption{Upper plot~: Variation of $E_n-\Re\cE_n^\one=E_n-\lambda_n\epsilon_\g-mg\Re(a)$ for antihydrogen interacting with a perfect mirror (blue top lines), a silicon bulk (green middle lines) or a silica bulk (red bottom lines), in units of $10^{-5}\epsilon_\g$. Points are obtained from numerical results $E_n^\num$ in \S\ref{sec:cavity} and full lines interpolate between these points. Dashed curves correspond to solutions $E_n^\ana$ of the effective-range equation \eqref{qsenergy}. Lower plot~: Difference $\Delta E_n$ between analytical and numerical energies on the upper plot for a perfect mirror (blue dotted line), a silicon bulk (green dashed line) or a silica bulk (red full line), in units of $10^{-6}\epsilon_\g$. [Colors online] }
\label{shift}
\end{figure}

In order to assess the precision of the results, we draw on the upper plot of Figure \ref{shift} the variation of $E_n-\lambda_n\epsilon_\g-mg\Re(a)$, that is also $E_n-\Re\cE_n^\one$, for the first quantum states of antihydrogen above a perfect mirror (blue top lines), a silicon bulk (green middle lines) or a silica bulk (red bottom lines). Points are obtained from the numerical results $E_n^\num$ discussed in the preceding section \S\ref{sec:cavity} for $n=1,2,\ldots,10$ with full lines interpolating between these points. Dashed lines are obtained from the solutions $E_n^\ana$ of the analytical effective-range equation (\ref{qsenergy}). For completeness, the differences between the analytical and numerical values are also plotted on the lower plot of Figure \ref{shift}
\begin{align}
\label{DeltaEn}
&\Delta E_n=E_n^\ana-E_n^\num~.
\end{align}
Figure \ref{shift} shows small oscillations of the numerical values around the smoother variation obtained from the analytical approximation. These oscillations remain smaller than a few $10^{-6}\epsilon_\g$ for the first ten quantum states, which means that the effective-range approximation is sufficient to compute the corrections caused by the CP interaction at this accuracy level.

\section{Complex CP shifts}
\label{sec:complex}

The round-trip factor $\rho$ is a causal scattering amplitude, that is also an analytic function of energy $E$. This function can be continued to the complex plane where the equation $\rho=1$ can now be solved for complex energies $\cE_n$, the imaginary part of which are related to the widths of the cavity resonances.

\begin{figure}[t]
\centering
\includegraphics[width=0.45\textwidth]{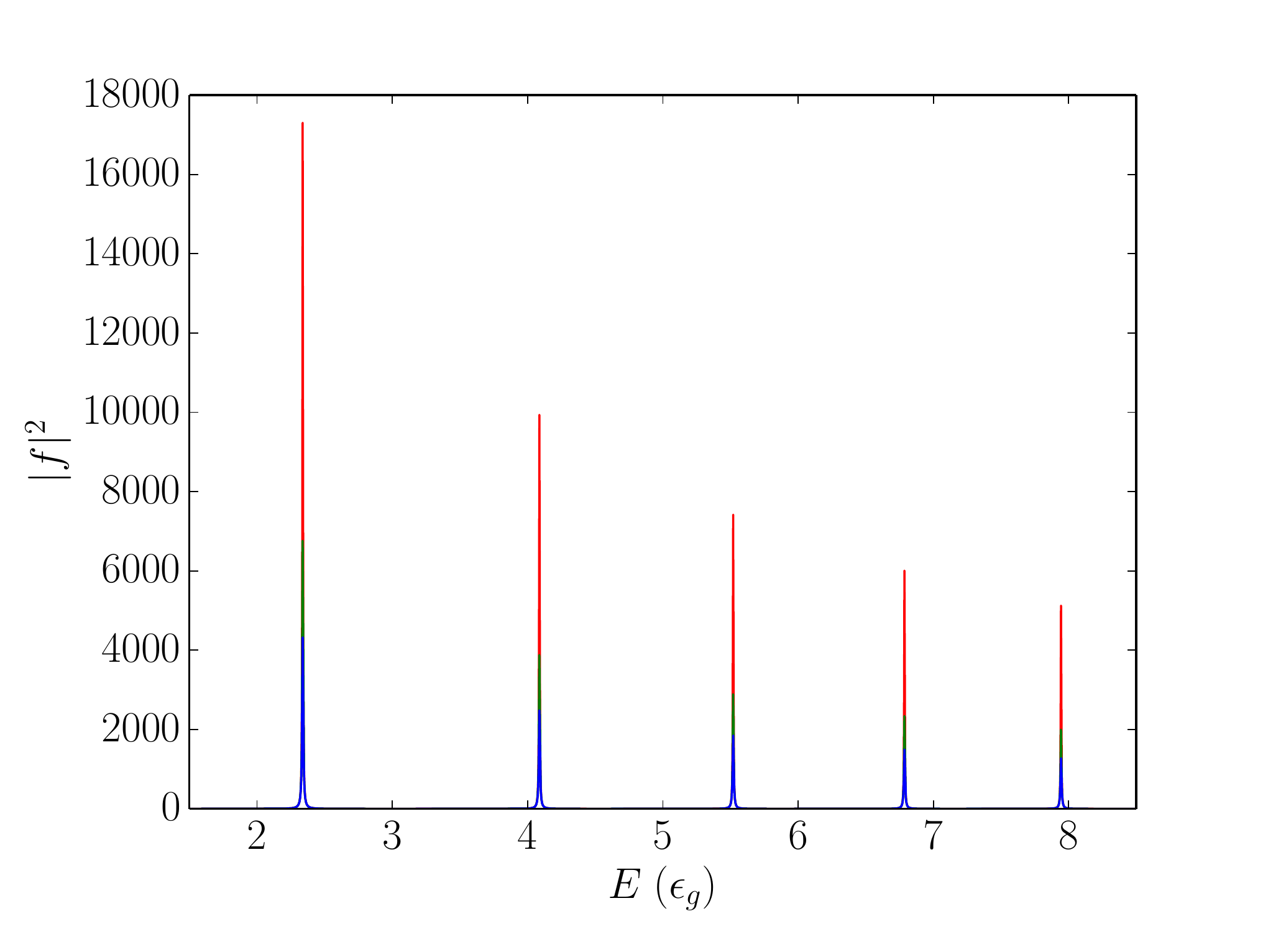}
\includegraphics[width=0.45\textwidth]{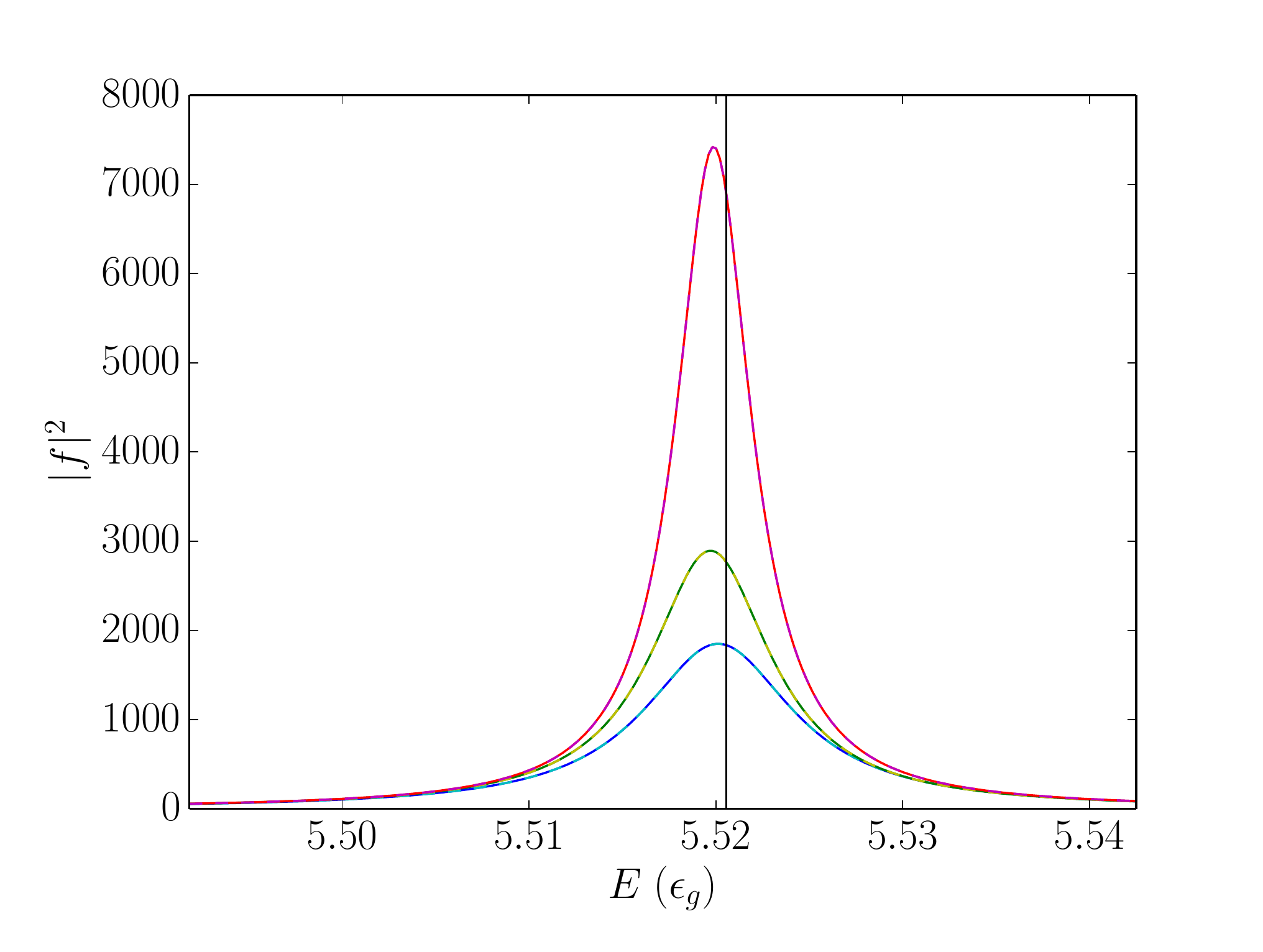}
\caption{\label{closed-loop} Upper plot : Squared modulus $|f(E)|^2$ of the closed loop function \eqref{closed-loop} as a function of energy for antihydrogen interacting with a perfect mirror (blue bottom lines), a silicon bulk (green middle lines) or a silica bulk (red top lines). Lower plot : Third peak (full line) and Lorentzian fit (dashed line) of the squared modulus $|f(E)|^2$ of the closed loop function for antihydrogen interacting with a perfect mirror (blue), a silicon bulk (green) or a silica bulk (red). The vertical line indicates the position of the ideal quantum bouncer energy $E_3^\zer=\lambda _3\epsilon_g$. [Colors online] }
\end{figure}

The complex solutions $\cE_n$ of the equation $\rho=1$ are also the poles of the \emph{cavity response function} accounting for multiple interference of different numbers of round trips  as for optical Fabry-Perot cavities \cite{Lambrecht2006}
\begin{align}\label{cavityfunction}
f\left(E\right)\equiv\frac{\rho\left(E\right)}{1-\rho\left(E\right)}=\rho+\rho^2+\rho^3+\ldots~.
\end{align}
When the reflection amplitude is replaced by its scattering length approximation, and the CP shifts treated perturbatively, these complex energies are obtained as $\cE_n^\one$ (see eq.\eqref{constant-shift}). The real part $\Re\cE_n$ is close to the resonance energy $E_n$ discussed in the preceding sections whereas the imaginary part $\Im\cE_n\simeq -mgb$ with $b=-\Im(a)$ is directly related to the width of the resonance and consequently to the inverse of the cavity lifetime.  In the scattering length approximation, the widths or lifetimes are thus determined by the same quantity $mgb$ for the different quantum states \cite{Voronin2005jpb,Voronin2005pra}. 

This is explained by a classical picture of the bounces. For an energy $E=mgH$, the bouncing period  is $2\sqrt{2H/g}$ while the probability of transmission through the quantum reflection barrier at each bounce is $1-|r|^2\simeq 4 \hbar^{-1}\sqrt{2 m E}b $. The lifetime is given by the ratio of these two quantities $\tau=\hbar/{(2 m g b)}$ which does not depend on $E$. The lower reflection probability at higher energies is compensated by the smaller bouncing frequency, so that the lifetime is independent of energy. This simple property is no longer exact with the more accurate treatment developed in the present paper.

\begin{figure}[t]
\centering
\includegraphics[width=0.45\textwidth]{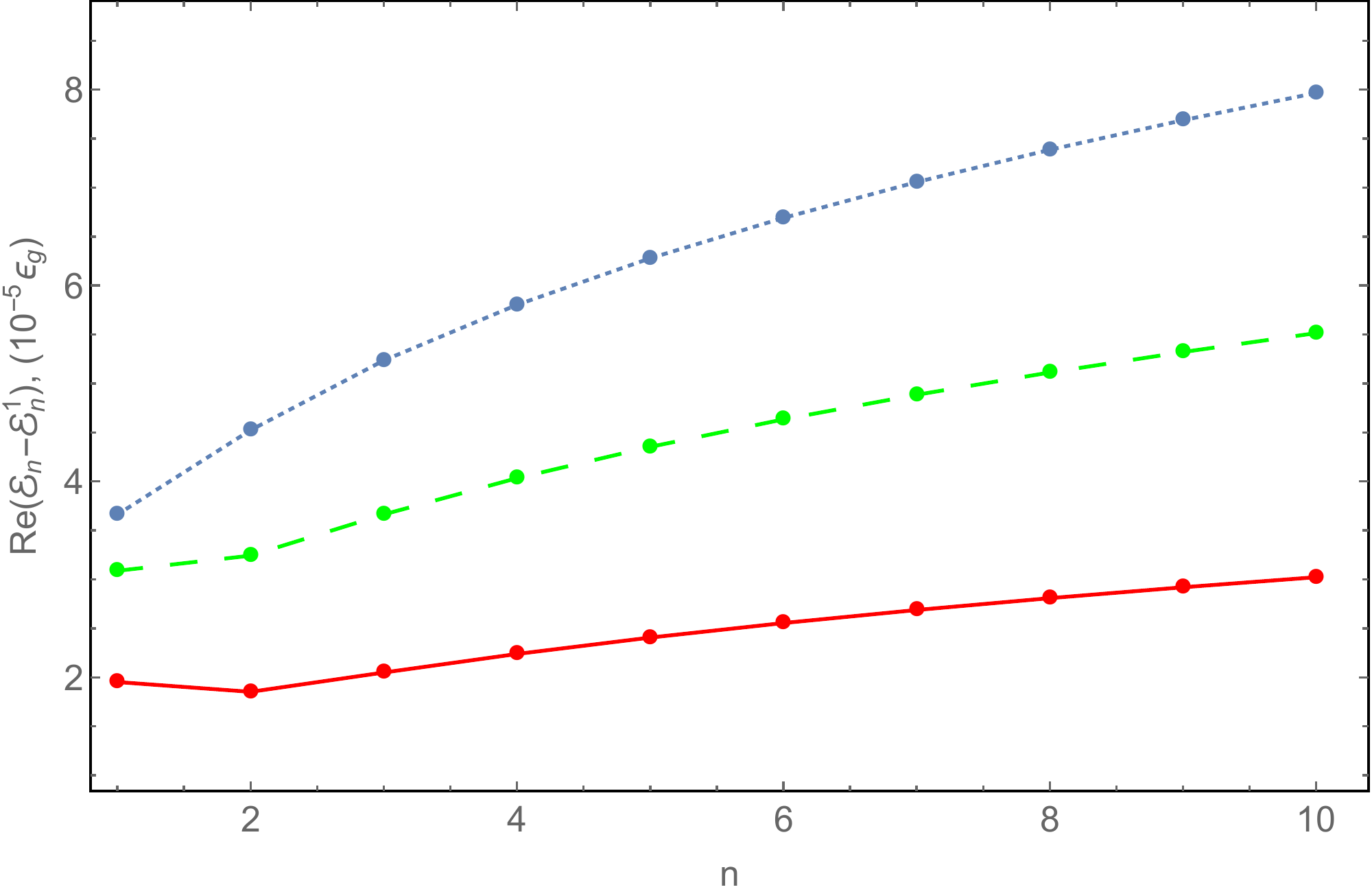}
\includegraphics[width=0.45\textwidth]{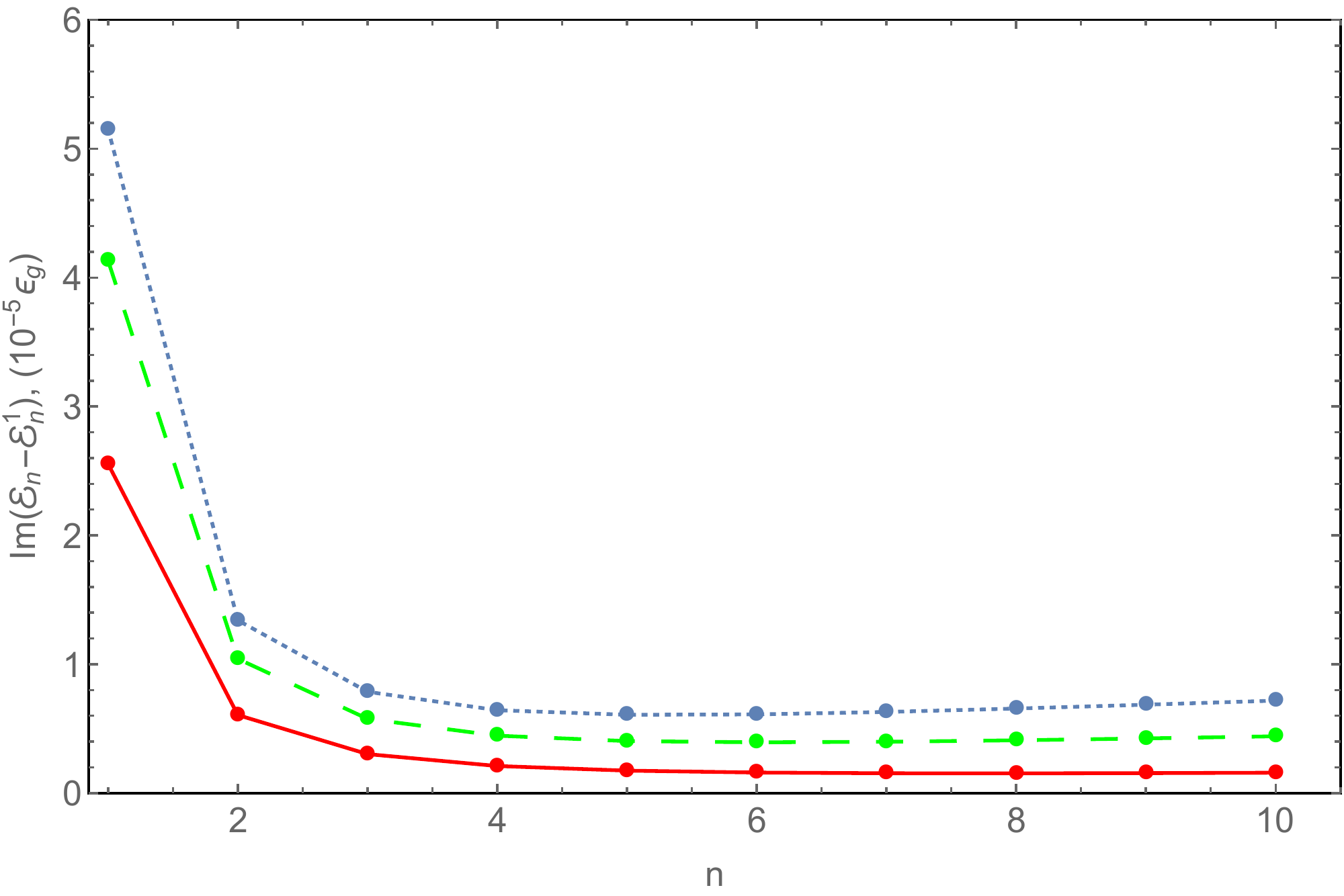}
\caption{Differences between the complex energies $\cE_n$ and their approximate expressions $\cE_n^\one=\lambda_n\epsilon_\g+mga$ for a perfect mirror (blue top line), a silicon bulk (green middle line) and a silica bulk (red bottom line). The upper and lower plots show the real and imaginary parts of these complex differences, both in units of $10^{-5}\epsilon_\g$. [Colors online] }
\label{Diff}
\end{figure}

In a first stage, we use the numerical results presented in \S\ref{sec:cavity} to obtain the complex energies. The upper plot on Fig.\ref{closed-loop} shows the first resonance peaks of $|f(E)|^2$ for antihydrogen interacting with a perfect mirror, a silicon bulk or a silica bulk.
The plot shows Lorentzian  resonances for $E$ close to the complex energies $\cE_n$
\begin{align}
\label{closed-loop-modsquared}
\left|f \right|^2 &  \simeq \frac{A_n}{\left|E-\cE_n\right|^2}=\frac{A_n}{\left(E-\Re\cE_n\right)^2+\left(\Im\cE_n\right)^2} ~, \\
&A_n= \frac{1}{\left|\rho_n^\prime \right|^2}~, \quad \rho_n^\prime\equiv\frac{\D\rho}{\D\cE}\left(\cE_n\right)~.
\end{align}
As the resonances are well separated, the contributions of other peaks have been disregarded in \eqref{closed-loop-modsquared}. The parameters for one peak can be retrieved by fitting numerical values $\left|f \right|^2 $ with \eqref{closed-loop-modsquared}. The lower plot on Fig.\ref{closed-loop} represents a zoom on the third peak which shows an excellent agreement with the fitting functions for the three different mirrors. 

We finally use the analytical approximate expressions results presented in \S\ref{sec:cpshifts} to obtain the complex energies. As the reflection amplitude and Airy phase function appearing in the expression \eqref{defrho} of $\rho$ are analytical functions, this is simply done by solving the equation $\rho\left(\cE\right)=1$ continued to the complex plane.
The results of these calculations are shown on Figure \ref{Diff} as differences between the complex energies $\cE_n$ and their scattering length approximations $\cE_n^\one=\lambda_n\epsilon_\g+mga$. The upper and lower plots correspond to real and imaginary parts of these differences and show that the differences are at a level of a few $10^{-5}\epsilon_\g$ for the lowest lying quantum levitation states.

\begin{figure}[t]
\centering
\includegraphics[width=0.45\textwidth]{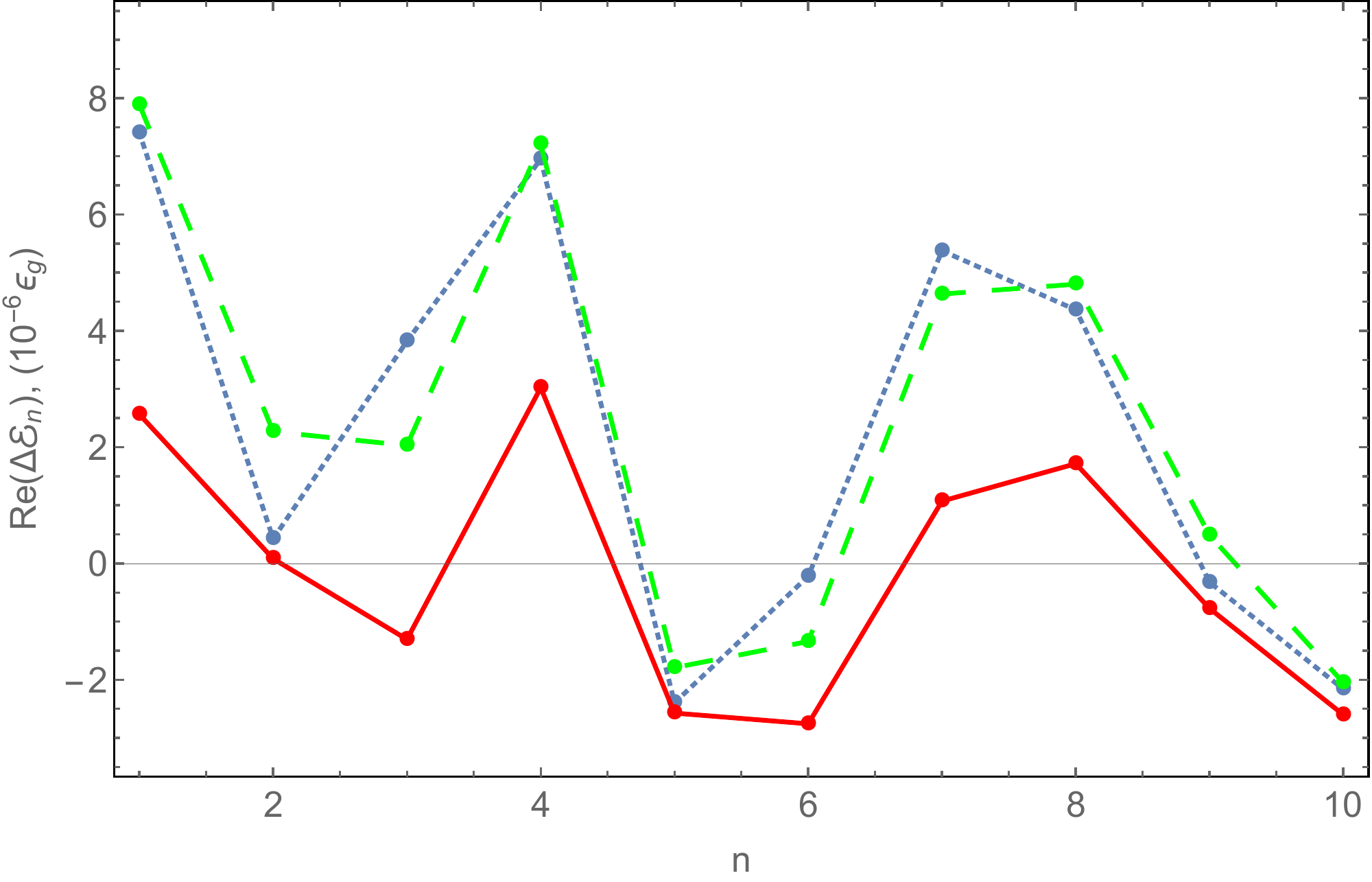}
\includegraphics[width=0.45\textwidth]{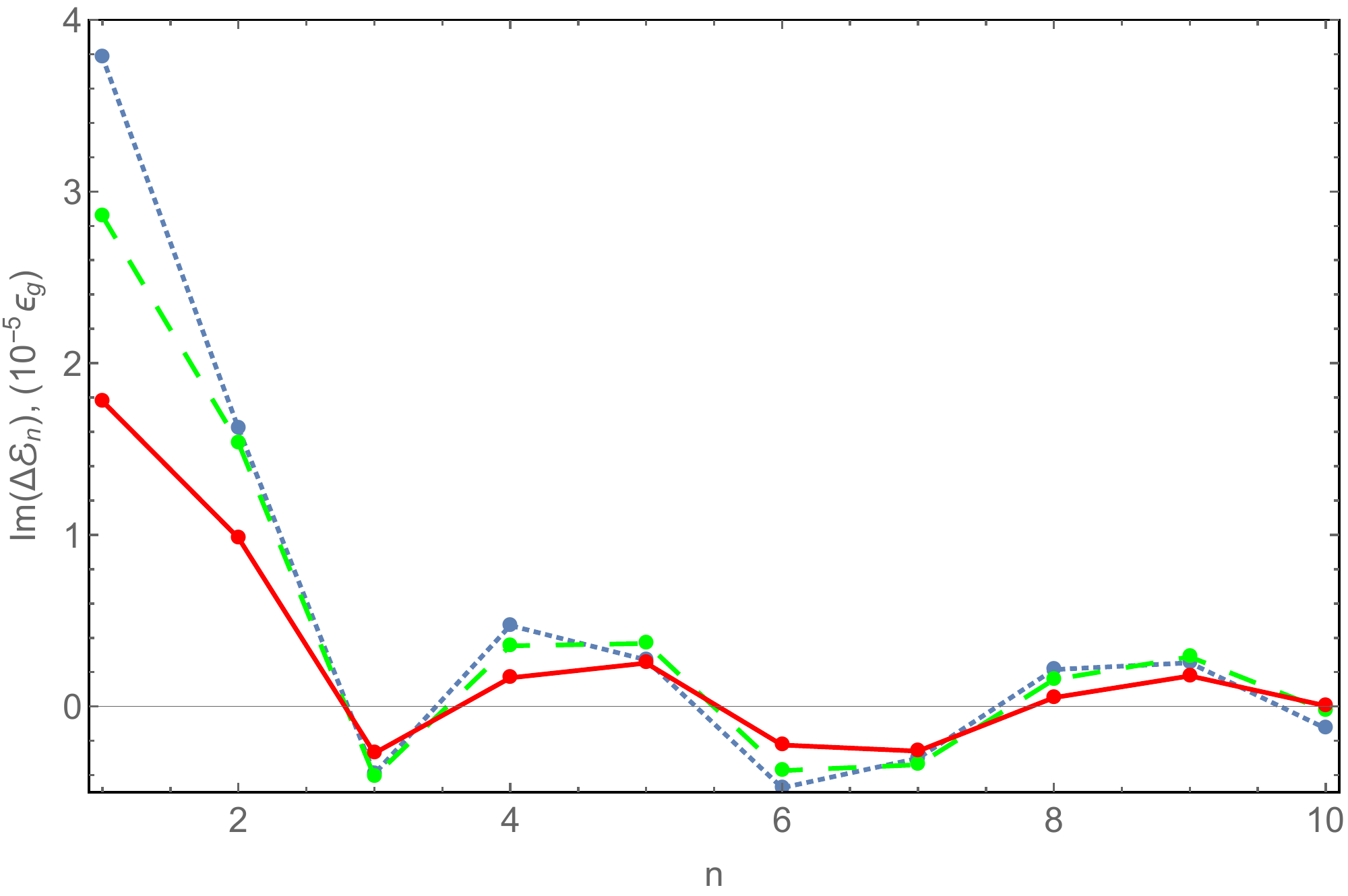}
\caption{Differences between the complex energies $\cE_n$ calculated in the analytical and numerical methods for a perfect mirror (blue dotted line), a silicon bulk (green dashed line) and a silica bulk (red full line). The upper and lower plots show the real and imaginary parts of these complex differences, in units of $10^{-6}\epsilon_\g$ and in units of $10^{-5}\epsilon_\g$ respectively. [Colors online] }
\label{DiffAnaNum}
\end{figure}

We show on Figure \ref{DiffAnaNum} the differences between the analytical and numerical solutions for complex energies
\begin{align}
\label{DeltacEn}
&\Delta \cE_n=\cE_n^\ana-\cE_n^\num ~.
\end{align}
The upper and lower plots show the real and imaginary parts of $\Delta \cE_n$, in units of $10^{-6}\epsilon_\g$ and $10^{-5}\epsilon_\g$ respectively. Both plots show oscillations of the numerical values around the smoother variation obtained from the analytical method. These oscillations remain at a level smaller than $8\times10^{-6}\epsilon_\g$ for the real part, $4\times10^{-5}\epsilon_\g$ for the imaginary part. Though the level of agreement is worse by a factor of the order of 5 for the imaginary parts than for the real ones, we note that the precise knowledge of resonance widths is less critical than that of resonance positions when analyzing spectroscopic measurements. 

In this paper, we have given detailed calculations of the Casimir-Polder shifts on quantum levitation states of antihydrogen atoms above a material surface. We have used Liouville transformations and Langer coordinates to build up a physical picture of these states corresponding to resonances of a cavity. The bottom mirror of the cavity is a partly reflective one associated with quantum reflection on the CP potential while the top mirror is a perfectly reflecting one due to gravity.  We have presented a full numerical treatment as well as an improved approximate analytical discussion of the properties of the cavity resonances. We have also proposed two different methods for characterizing these properties through the complex energies $\cE_n$ defined as poles of the cavity response function (calculations in \S\ref{sec:complex}) or the real resonance energies $E_n$  defined from the cavity round trip phase (calculations in \S\ref{sec:cpshifts}). We have checked that $\left|\Re\cE_n-E_n\right|<3\times10^{-6}\epsilon_\g$ for all low-lying quantum states. 

The comparison of all these results shows that the analytical treatment built up in the present paper on the effective-range approximation is sufficient to compute the corrections caused by the CP interaction at an accuracy level better than $10^{-5}\epsilon_\g$ for the positions of the resonances. This should be sufficient for analyzing spectroscopic tests  of the weak equivalence principle with antihydrogen \cite{Voronin2011,Voronin2014ijmp,Voronin2016jpb} up to an accuracy of this order.

\textit{Acknowledgements - } 
{Thanks are due to M.-T. Jaekel, V.V. Nesvizhevsky, A. Yu. Voronin for insightful discussions and to the GBAR and GRANIT
collaborations. }


%

\end{document}